\documentclass[aps,prb,twocolumn,floats,superscriptaddress]{revtex4}
\usepackage{epsfig}
\usepackage{amsmath}
\usepackage{amssymb}

\newcommand{\beq}{\begin{equation}}
\newcommand{\eeq}{\end{equation}}
\newcommand{\bea}{\begin{eqnarray}}
\newcommand{\eea}{\end{eqnarray}}

\newcommand{\eps}{\epsilon}
\newcommand{\veps}{\varepsilon}
\newcommand{\al}{\alpha}
\newcommand{\s}{\sigma}

\newcommand{\om}{\omega}

\newcommand{\be}{\beta}

\newcommand{\vp}{\varphi}

\newcommand{\ua}{\uparrow}
\newcommand{\da}{\downarrow}

\newcommand{\w}{\omega}
\newcommand{\pdag}{{\phantom{\dagger}}}

\newcommand{\intx}{\int_{-\infty}^{+\infty}\!\!\!\!\!\!\!\!\!\! dx}

\begin{document}
\title{Interplay of electromagnetic noise and Kondo effect in quantum dots
}
\author{Serge Florens}
\affiliation{Institut N\'eel, Centre National de la Recherche Scientifique and Universit\'e Joseph Fourier, BP 166, 38042 Grenoble, France}
\affiliation{Institut f\"ur Theorie der Kondensierten Materie,
Universit\"at Karlsruhe, 76128 Karlsruhe, Germany}
\author{Pascal Simon}
\affiliation{Laboratoire de Physique et Mod\'elisation des Milieux
Condens\'es, Centre National de la Recherche Scientifique and Universit\'e Joseph Fourier, 38042 Grenoble, France}
\affiliation{Department of Physics and Astronomy, University of Basel,
Klingelbergstrasse 82, CH-4056, Basel, Switzerland}
\author{Sabine Andergassen}
\affiliation{Institut N\'eel, Centre National de la Recherche Scientifique and Universit\'e Joseph Fourier, BP 166, 38042 Grenoble, France}
\author{Denis Feinberg}
\affiliation{Institut N\'eel, Centre National de la Recherche Scientifique and Universit\'e Joseph Fourier, BP 166, 38042 Grenoble, France}

\begin{abstract}
We investigate the influence of an electromagnetic environment, characterized 
by a finite impedance $Z(\om)$, on the Kondo effect in quantum dots.
The circuit voltage fluctuations couple to charge fluctuations in the dot
and  influence the spin exchange processes transferring charge between the
electrodes. We discuss how the low-energy properties of a Kondo quantum dot 
subject to dynamical Coulomb blockade resemble those of Kondo impurities in
Luttinger liquids. Using previous knowledge based on the bosonization of 
quantum impurity models, we show that low-voltage conductance anomalies appear 
at zero temperature. The conductance can vanish  at low temperatures even in the 
presence of a screened impurity spin. Moreover, the quantitative determination 
of the corresponding Kondo temperature depends on the full frequency-dependent 
impedance of the circuit. This is demonstrated by a weak-coupling calculation 
in the Kondo interaction, taking into account the full distribution $P(E)$ 
of excited environmental modes.
\end{abstract}

\pacs{72.15.Qm,71.10.Pm,72.10.Fk,73.63.Kv}

\maketitle


\section{Introduction}

Recent progress in controlling the electronic properties of
semiconductor-based nanostructures provides new ways of probing the physics of
strong correlations, and envisions a rich interplay with truly mesoscopic
effects. One of the most prominent examples 
is certainly the discovery of the Kondo effect in quantum
dots.\cite{goldhaber,revival_kondo} While in metals containing a small amount
of magnetic impurities the increased magnetic scattering of the electrons at
low temperature results in an increased resistance, the mesoscopic realization
displays instead a zero-bias peak in the conductance.\cite{glazmanraikh} This
setup opens also a way for the study of various mesoscopic phenomena in the
presence of strong correlations, such as nonequilibrium transport
regimes~\cite{kogan,winmeir,kaminski} and finite-size effects in the
electrodes.\cite{simon02,finite_size}

One should emphasize that the unitary conductance, a feature in the charge
sector, is currently taken as the fingerprint of the Kondo effect, a property of
the impurity spin being screened by the conduction electron spins. 
Yet, it is not obvious that unitary conductance and spin screening should always coincide. 
In fact, one should bear in mind that conductance through a single or a double junction is
sensitive to the circuit environment, made for instance of a resistance $R$. The parameter
$r=R/R_K$ defines the normalized resistance with $R_K=h/e^2$. For
such an ohmic bath, in a strongly resistive single junction of resistance $R_t >> R_K$ and
capacitance $C$, with charging energy $E_c=e^2/2C$, a low-bias anomaly $dI/dV
\sim (V/E_c)^{2r}$ appears. This phenomenon is called dynamical Coulomb blockade
(DCB), (see Refs.~[\onlinecite{devgrab,IN}] for a review). It has been extensively
studied in normal single tunnel junctions with resistive leads both
experimentally~\cite{delsing,cleland,joyez_esteve_devoret} and
theoretically,\cite{nazarov,girvin,odintsov_single,devoret,flensberg_jonson,panyukov,wang,safi,levy-yeyati}
as well as for general scatterers \cite{Nazarov99,golubev}, in normal double tunnel
junctions,\cite{odintsov_double,joyez_esteve} superconducting
junctions~\cite{schon,watanabe,rimberg} and single barriers in semiconductors.\cite{popovic}
On the other hand, for transparent junctions with $R_t < R_K$, the same
phenomenon occurs, but only below an exponentially small DCB  energy scale
$\tilde{E}_C \sim E_c e^{-R_K/2R_t}$. Only when at least one channel becomes
fully transparent,  dynamical Coulomb blockade disappears and a linear $I(V)$ is
recovered \cite{panyukov,FlensbergLL,matveev,Nazarov99,golubev}. Coming back to the Kondo
problem, one readily sees the difficulty encountered in trying to guess the
low-temperature behavior. On one hand, the zero-temperature regime being fully
transparent, one might think that the Kondo effect - and the unitary conductance
- are not affected by an ohmic bath. Yet, at any finite temperature, the
 transmission being less than one, DCB should reappear. In other words, it is
not clear whether the Kondo temperature is lower or higher than a DCB energy scale.
This problem obviously cannot be solved without reconsidering how the couplings
leading to the Kondo effect (and the unitary-conductance peak) renormalize at
low energy in presence of environmental fluctuations. This is the issue we address
in the present paper.

Previous investigations of the Kondo effect with  dynamical Coulomb blockade
focused on the regime where the quantum dot is near the charge degeneracy point,
with a noisy back gate directly coupled to the quantum dot.\cite{karyn1,karyn2,BZS,BZGG}
In this case, a Kondo model for the charge degree of freedom can be derived, with 
a direct coupling of the dissipation to this charge variable. On general grounds 
this leads to a competition between the Kondo screening of the charge doublet 
by the electrons and the localization effect due to the ohmic environment.

Here, we concentrate rather on the usual spin Kondo effect,
with environmental electromagnetic fluctuations in the electrodes. In this case, 
dissipation induces a markedly different effect on the Kondo physics. Indeed, 
only the inter-lead spin exchange processes involve charge fluctuations across 
the device and hence dominantly couple to the environment.
One then assists to a competition between several Kondo-type strong-coupling fixed 
points in absence of localization effects. Different conduction regimes may then 
be approached, depending on the dissipation strength and microscopic couplings.

At weak dissipation, one expects to preserve the usual Kondo effect, but 
dephasing of the couplings involving charge transfer affects the determination 
of the Kondo temperature $T_K$ below which the spin is 
eventually screened on a quantitative level. For this purpose we
examine how the perturbative regime of the Kondo model is modified by
a general dissipative environment in the formalism of the $P(E)$ theory.
Focusing on a circuit with a finite zero-frequency impedance $Z(\w\!=\!0)=R$, 
but keeping a complete description of the environment characterized by the
above low-energy ohmic behavior and a higher energy tail related to the circuit 
capacitances, we explicitely calculate how those parameters influence the 
temperature scale $T_K$ at which the Kondo resonance forms. At temperatures lower 
than $T_K$, we can apply previous knowledge~\cite{kane,fabrizio,kim,simon01,AEM} on the Kondo effect 
between Luttinger liquids (to which our setup is equivalent in some low-energy limit {\it only})
to capture the current-voltage relation. This extends the previous mappings  
for single or double junctions~\cite{FlensbergLL,safi}.
For the weak-dissipation regime, we thus find that nonohmic behavior, thus DCB, prevails
in presence of particle-hole asymmetry. Particle-hole symmetry can be restored for
some specific values of the dot gate voltage allowing for a unitary conductance,
albeit with anomalous corrections.

At large dissipation, which may be more challenging to realize experimentally,
the electron transfer processes are even more strongly suppressed. As soon as the 
environmental impedance $R$ reaches half the quantum value $h/e^2$, a generic nonohmic 
transport regime develops (valid independently of particle-hole symmetry, but with 
unbalanced left/right couplings), where the Kondo effect occurs through the strongest coupled
electrode only. With balanced couplings, a two-channel Kondo regime may be feasible,
and represents a generalization of a recently proposed ~\cite{oreggold,SFAR} and experimentally studied setup \cite{goldhaber06} 
with strong Coulomb blockaded leads, formally realizing the limit $r\gg1$. 
We note that our proposal does not suffer from a low-energy cutoff set by the level 
spacing in the leads. Notice that weak dissipation ($R\ll R_K$) is not sufficient to reach such
non-Fermi liquid fixed point.

The paper is organized as follows. In Sec.~\ref{sec:dynCB} we give a characterization
of the ohmic environment, and introduce the various physically relevant scales 
for the problem. In Sec.~\ref{sec:spinkondo} we propose a 
phenomenological model describing how the Kondo effect may be
affected by those environmental fluctuations. This model is analyzed both at
weak coupling, taking into account the full spectral function $P(E)$ of the
environment, and at strong coupling (zero temperature), using a low-energy mapping 
onto a Luttinger chain through bosonization, explicitly derived in Appendix~\ref{app}. 
We conclude the paper with an outlook on possible future theoretical developments and
experimental signatures of  dynamical Coulomb blockade on the Kondo physics.

\section{Environmental Coulomb blockade}
\label{sec:dynCB}

\subsection{Circuit theory}
In this part we briefly summarize the basics of circuit theory.\cite{IN,devgrab} 
This enables us to introduce also some notations
that will be used along the paper. The device we are interested in consists of
a nano-object, which can be an artificial atom, a molecule or another interacting 
system, connected to reservoirs by two tunnel junctions. Let us call S this interacting system. Each
time an electron tunnels in or out the system S, this electron can excite modes
in the electromagnetic environment, describing the circuit external to the
tunnel junction. As the environment comprises S itself, the crucial part
is to properly describe the tunnel junctions. A semi-phenomenological
approach able to capture these effects has been developed by Nazarov~\cite{nazarov,IN} 
and Devoret {\it et al.}.\cite{devoret} The environment is modeled
by its own impedance $Z(\omega)$ (now arbitrary), leading to an impedance 
$Z_t(\omega)$ seen by the junction. 
We denote by $Q$ the charge displacement at the surface of the capacitance caused by a
tunneling event. Note that $Q$ is a collective variable. We also define $\vp$
the phase conjugate to $Q$ satisfying $[\vp,Q]=ie$. Let us now derive the correlation
functions of the phase variable $\vp$ in the case of ohmic fluctuations.

\subsection{Ohmic phase fluctuation spectrum}
Working with imaginary frequencies, we can parametrize the bosonic Green
function~\cite{caldeira} for the phase variable as
\begin{eqnarray}
G_\vp(i\w) & \equiv & \langle\vp(i\w)\vp(-i\w)\rangle\\
\label{spectrum}
& = & 2\pi \left[\frac{R_K}{Z(i\w)} |\w| + R_K C \w^2\right]^{-1}\;,
\end{eqnarray}
where the first contribution in Eq.~(\ref{spectrum}) defines the the 
frequency-dependent impedance $Z(\w)$ of the circuit,\cite{IN} 
while the second is associated with the junction capacitance $C$
(see Sec.~\ref{model} for details). We take $\hbar=1$ in the following. 
For an ohmic environment, the circuit resistance is given by $R=Z(\w\!=\!0)$,
which from Eq.~(\ref{spectrum}) yields the natural cutoff frequency $\w_R=1/(RC)$.
However, even for strong dissipation ($R\sim R_K$) $\w_R$ is a very large frequency,
at which $Z(\w)$ cannot be reduced to its zero-frequency value. The
circuit impedance typically decays as $Z(\w) \sim \sqrt{R/|\w| C_l}=R\sqrt{\w_l/\w}$
for $\w$ larger than $\w_l$; $\w_l=1/(RC_l)$, and $C_l$ is a measure of the
leads capacitance.\cite{devgrab} To simplify further calculations, we will
model~\cite{cron} this tail of the circuit impedance $Z(\w)$ by a pure capacitive 
decay $Z(\w) \sim R\w_l/\w$. This allows to rewrite the phase correlations as
\begin{equation}
\label{altspectrum}
G_\vp(i\w) = 2\pi \frac{R}{R_K} \left[ |\w| + \frac{\w^2}{\w_c}\right]^{-1}\;,
\end{equation}
where $\w_c = \w_l\w_R/(\w_l+\w_R)$ is an effective cutoff. We introduce also 
the related effective capacitance $C_{\rm eff}$ controlling the high-frequency tail of the
phase fluctuations $\w_c=1/(RC_{\rm eff})$. Using simple 
complex analysis, we can extract the real-frequency spectral function
\begin{equation}
\mathcal{I}m \; G_\vp(\w) 
= 2\pi \frac{R}{R_K} \left[ \w + \frac{\w^3}{\w_c^2}\right]^{-1}
\label{realspectrum}
\end{equation}
which obeys
\begin{equation}
G_\vp(i\w) = - \int \frac{d\epsilon}{\pi} 
\frac{\mathcal{I}m [G_\vp(\epsilon)]}{i\w-\epsilon}\;.
\end{equation}
Fourier transforming in imaginary time $\tau$ yields
\begin{equation}
G_\vp(\tau) = - \int \frac{d\epsilon}{\pi} \frac{1}{e^{-\beta\epsilon}-1}
\mathcal{I}m [G_\vp(\epsilon)] e^{-\epsilon\tau}\;,
\end{equation}
where $\beta=1/(k_BT)$ is the inverse temperature, and performing the analytic
continuation $\tau\rightarrow it$ to real time $t$, we get the usual formula~\cite{IN}
\begin{eqnarray}
J_\vp(t) & \equiv & \langle(\vp(t)-\vp(0))\vp(0)\rangle\\
\nonumber
& = & \int\limits_0^\infty\frac{d\omega}{\omega}
\frac{2\mathcal{R}e[Z_t(\omega)]}{R_K}\left(\frac{\cos(\omega t)-1}{\tanh(\beta\omega/2)}
-i\sin(\omega t)\right)\;,
\end{eqnarray}
where $Z_t(\w) = [i\w C+1/R]^{-1}$ is the impedance seen by the junction.
Since charge transfer processes are associated to the operator $e^{i\vp}$ (see
Sec.~\ref{model}), it is useful to consider the real-time correlation function
$\langle e^{i\vp(t)-i\vp(0)}\rangle = e^{J_\vp(t)}$, which by Fourier transformation 
\beq
P(E)=\frac{1}{2\pi}\int_{-\infty}^{\infty} dt \; e^{J_\vp(t)+iEt}
\eeq
describes the probability $P(E)$ that the electron exchanges an energy $E$ with the 
environment during the tunneling process. 

At zero temperature and for long time
\beq
\label{asymptot}
J_\vp(t)\sim -2r\ln(i\om_c t)\;,
\eeq
where $r=R/R_K$ is the dimensionless resistance of the circuit, which leads to
the low-energy behavior~\cite{IN}
\beq
\label{low}
P(E) \sim A_r \frac{E^{2r-1}}{\w_c^{2r}}\hspace{.5cm}\rm{for}\quad E\ll\w_c,
\eeq
with $A_r = e^{-2 r\gamma}/\Gamma(2r)$.
This power law is however cut off for energies larger than $\w_c$; in this range 
the probability decays as
\beq
\label{high}
P(E) \sim 2r \frac{\w_c^2}{E^3}\hspace{.5cm}\rm{for}\quad E\gg\w_c \;.
\eeq
We finally mention some useful properties~\cite{IN} of the $P(E)$ function:
i) detailed balance $P(E)=0$ for $E<0$ at $T=0$; ii) normalization
$\int_0^{+\infty} dE \; P(E) = 1$; iii) integral equation
\beq
\label{inteq}
E P(E) = 2r \int_0^E d\w \; \frac{P(\w)}{1+\left(\frac{\w-E}{\w_c}\right)^2}\;.
\eeq 

In the next section we analyze how the Kondo effect occurring in 
mesoscopic quantum dots is modified by the presence of such an ohmic environment.

\section{The noisy Kondo effect}
\label{sec:spinkondo}

\subsection{Derivation of the model}
\label{model}

\begin{figure}[ht]
\epsfig{figure=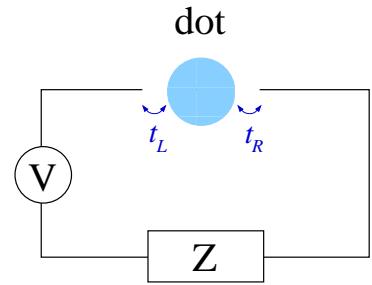,width=4.75cm}
\caption{(Color online) Schematic representation of a circuit with a dot coupled to two leads 
and an impedance $Z$.
}\label{Fig:circuit1}
\end{figure}

In this section, we consider a single quantum dot  connected to
two metallic leads in the Kondo regime, and investigate how the Kondo effect
is affected by the presence of environmental Coulomb blockade.
We assume that electromagnetic fluctuations are modeled by an impedance $Z(\omega)\sim R$
closing the circuit as described above.
The equivalent circuit depicted in Fig.~\ref{Fig:circuit1} is the one of a single-electron 
transistor.

A model Hamiltonian describing this situation can be written as 
$H=H_{\rm dot}+H_{\rm leads}+H_{\rm tun}+H_{\rm bath}$,
where
\bea 
\label{dot}
H_{\rm dot}&=&\sum_\s\eps_d d_{\s}^\dag d_\s^\pdag+ Un_{d,\ua}n_{d,\da},\\
\label{leads}
H_{\rm leads}&=&\sum\limits_{k,\s,\al=L/R}\eps_k c^\dag_{k,\s,\al}c^\pdag_{k,\s,\al},\\
\nonumber
\label{tun}
H_{\rm tun}&=&\sum\limits_{k,\s} (t_L e^{i\vp_L} d^\dag_{\s} c^\pdag_{k,\s,L}
+t_R e^{i\vp_R}c^\dag_{k,\s,R}d^\pdag_{\s})+\rm{H.c.}.\\
\eea
$H_{\rm bath}$ describes the environmental degrees of freedom consisting
of a collection of harmonic oscillators, alternatively expressed by a phase
correlator such as Eq.~(\ref{spectrum}).
In these equations, $c^\dag_{k,\s,\al}$ creates an electron with energy
$\epsilon_{k}$ in lead $\al =L,R$ with spin $\s$, and $d^\dag_{\s}$ creates an
electron in the dot with spin $\s$. $H_{\rm dot}$ is the usual Anderson
Hamiltonian, where $\epsilon_d$ is the dot energy controlled by a gate voltage
$V_g$, while $U$ describes the Coulomb interaction within the dot. The phases
$\vp_\al$ appearing in the tunneling Hamiltonian are related to the voltage
fluctuations $\delta V_\al$ felt by an electron during a tunneling event through
the $L/R$ junction as follows
\beq
\vp_\al(t)=e \int\limits_{-\infty}^t \delta V_\al(t') dt'\;.
\eeq
The phases $\vp_\al$ are conjugate to the charge $Q_\al$ on the junction
capacitance $\al$ such that $[\vp_\al,Q_\al]=ie$. As they originate from the
same bath, the phases $\vp_\al$ are clearly not independent. In order to relate
the voltage fluctuations in the junction $\al$ to $Z(\om)$, we use the standard
effective-circuit description:\cite{IN} when an electron tunnels in
the left junction, it feels an environment that can be modeled by the right
junction in parallel with $Z(\w)$. The left/right junctions are simply modeled by a
tunneling resistance $R_t$ in parallel with a capacitance $C_L$/$C_R$. This
description is consistent with the single-electron tunneling terms in Eq.~(\ref{tun}), 
and may be derived more microscopically by coupling independently each of the tunneling 
operators to the bosonic charge modes 
in the circuit~\cite{Imam_Averin,odintsov_double}. Neglecting the Kondo correlations for a moment, the
setup of Fig.~\ref{Fig:circuit1} is simply a double junction, which has been
treated in great detail in Ref.~[\onlinecite{IN}] in the sequential regime. For a
quantum mechanical description of the double junction in series, it turns out to be more
convenient to work with the charges $q=Q_L-Q_R$ and
$Q=\frac{C_RQ_L+C_LQ_R}{C_L+C_R}$ being uncoupled in the absence of
tunneling; $q$ corresponds to the dot charge quantized in units of $e$ and $Q$
is simply the total charge carried by the capacitance $C=\frac{C_LC_R}{C_L+C_R}$
that couples directly to the impedance $Z$.
One may then introduce the phases $\vp$ and $\psi$ conjugate to $Q$
and $q$ respectively, related to the phases $\vp_\al$ by
$\vp_L=(C/C_L)\vp+\psi$ and $\vp_R=-(C/C_R)\vp+\psi$. In the equivalent circuit,
we neglected the part corresponding to the gate voltage controlling the
number of electrons on the dot. This approximation is justified, as
the capacitance $C_g$ associated to the gate in general satisfies $C_g\ll C_\al$ and
gate voltage fluctuations usually play a minor role 
(their contribution has been addressed in Ref.~\onlinecite{Ingold-gate}).

Since we are interested in the Kondo regime, where the real charge fluctuations
in the dot are suppressed, it is convenient to perform a generalized
Schrieffer-Wolff transformation, or equivalently integrate out the dot charge
degrees of freedom. The complication due to the time-dependence of the
tunneling term in Eq.~(\ref{tun}) can be circumvented by a few additional
assumptions. First we will assume that $|\delta V|\ll |\veps_d|,U+\veps_d$ which
constitutes no serious limitation. We further suppose that the energy
excitations of the environmental bosonic modes $\omega$ are small enough
to satisfy $ \omega\ll |\veps_d|,U+\veps_d$. Under these two conditions, the
dot is modeled by a spin $\vec S$ described by a Kondo-like Hamiltonian $H_K$, 
involving phase-dependent couplings
\beq
\label{HK}
H_K=\sum_{\al,\be} J_{\al,\be} \; e^{i(\vp_\al-\vp_\beta)} \! 
\sum\limits_{k,\s,k'\s'} c^\dag_{k,\s,\al}
\frac{\vec \tau_{\s,\s'}}{2} c^\pdag_{k',\s',\be} \cdot \vec S \;,
\eeq
where 
\beq
\label{JK}
J_{\al,\beta}\approx 2 t_\al t_\be \left(\frac{-1}{\veps_d}+
\frac{1}{U+\veps_d}\right)\;.
\eeq
In Eq.~(\ref{HK}) we should {\it a priori} include a dressed potential scattering term 
\beq 
\label{scat}
H_V=\sum_{\al,\be} V_{\al,\be} e^{i(\vp_\al-\vp_\beta)} \sum\limits_{k,\s,k'\s'} c^\dag_{k,\s,\al}c^\pdag_{k',\s',\be},
\eeq
where $V_{\al,\beta}\approx 2 t_\al t_\be \left(\frac{-1}{\veps_d}-
\frac{1}{U+\veps_d}\right)$. Nevertheless, the coupling constants $V_{\al\be}$
turn out to be either irrelevant or marginal. As they are not modifying our
weak-coupling analysis, we neglect these terms. However, these terms associated to particle-hole 
symmetry breaking are important near the strong-coupling fixed point as we will see.

The omission of the environment  fluctuations with respect to $\eps_d$ or $U$ in Eq.~(\ref{JK})
means that during a virtual charge fluctuation or cotunneling event~\cite{averin-nazarov} 
no energy is exchanged with the environment. 
This assumption is justified by the low value of the circuit cutoff frequency $\omega_l$. 
It contrasts for instance with previous treatments of so-called 
inelastic cotunneling~\cite{odintsov_double}, or coupling with local phonon modes~\cite{flensberg}.
We expect that going beyond this approximation would lead to small corrections to the model defined above, and their treatment (see Ref.~[\onlinecite{flensberg}] for a different setup and simpler
environment) is clearly beyond the scope of the present paper. Therefore, the elementary Kondo couplings result from quasi-"elastic" cotunneling (in the sense of Ref.~[\onlinecite{averin-nazarov}]),
where the initial and final states of the spin-flip transition may however involve different
environmental energies. 
The inter-lead or "backscattering" Kondo couplings $J_{LR}$ and $J_{RL}$ now include the effects of 
the environment, embodied by the dynamical phase $\vp_L(t)-\vp_R(t)=\vp(t)$. 
This dynamical phase clearly modifies the behavior of these couplings, 
which we will analyze in the next two subsections. On the contrary, within our quasi-elastic approximation, the 
intra-lead couplings $J_{LL}$ and $J_{RR}$ are not dressed by phase fluctuations.

\subsection{Weak Kondo coupling analysis}
\subsubsection{Poor-man's scaling}

The Hamiltonian~(\ref{HK}) can be regarded as the usual Kondo model, where
the inter-lead spin processes $J_{\al,\be}$, with $\al\neq\be$
transferring charge between the electrodes, are dressed by fluctuating phases 
described by the ohmic spectrum of Eq.~(\ref{spectrum}). The case of strong Coulomb blockade, 
corresponding to $R=\infty$, has been previously studied in Ref.~[\onlinecite{SFAR}]. The calculation 
of observables to lowest order amounts to compute the diagrammatics at order
$J^2$. Clearly the renormalization of the $J_{LR}$ vertex depends on the
combination $J_{LR} (J_{LL}+J_{RR})$, such that a single unpaired $e^{i\vp}$ term 
occurs, and does not affect the usual renormalization equation
\begin{equation}
J_{LR}' = J_{LR} + J_{LR} (J_{LL}+J_{RR}) \int_{\Lambda-\delta\Lambda}^\Lambda
d\epsilon\; \frac{\rho_0(\epsilon)} {-\epsilon}\;,
\end{equation}
where $\rho_0(\epsilon) = 1/2W$ is the free electron density of states and 
$W$ is the electronic bandwidth.
Following the poor-man's scaling philosophy, we introduce here a small shift 
$\delta \Lambda$ in the running cutoff $\Lambda$ and computed the resulting change in the
Kondo coupling. This flow starts at some high-energy initial cutoff $D \sim \rm{min}(U,W)$,
and reduces towards lower energies as long as it is perturbatively controlled. 
Introducing dimensionless couplings $j_{\al,\be}\equiv\rho_0 J_{\al,\be}$ yields
\begin{equation}
\label{flow1}
\frac{d j_{LR}}{d\log \Lambda} = - j_{LR}(j_{LL}+j_{RR})\;.
\end{equation}

On the contrary, the $J_{LL}$ process is obtained from the combination
$J_{LL}^2+J_{LR}^2$, so that the second term 
involves the combination of two
exponential phase factors. This amounts to replace in the diagrammatics the free
electronic Green function $G_0(\tau) = \langle c_R^\dag(\tau) c_{R}^\pdag(0)
\rangle$ by a mixed one:
\begin{eqnarray}
G_{mix}(\tau) & = & \langle c_R^\dag(\tau) c_{R}^\pdag(0) e^{i\vp(\tau)}
e^{-i\vp(0)}\rangle \\
& \approx & G_0(\tau) e^{J_\vp(\tau)}\;.
\end{eqnarray}

We denote $\rho_{mix}$ its associated density of states. Performing the analytic
continuation of the above Green function at zero temperature, we find the simple
formula 
\beq
\rho_{mix}(\w) = \rho_0 \int_0^{|\w|} dE \; P(E).
\eeq
The renormalization equation thus reads
\begin{equation}
J_{LL}' = J_{LL} + J_{LL}^2 \int_{\Lambda-\delta\Lambda}^\Lambda
d\epsilon\; \frac{\rho_{0}(\epsilon)} {-\epsilon} 
+ J_{LR}^2 \int_{\Lambda-\delta\Lambda}^\Lambda
d\epsilon\; \frac{\rho_{mix}(\epsilon)} {-\epsilon}
\end{equation}
leading to the flow equation
\begin{equation}
\label{flow2}
\frac{d j_{LL}}{d\log \Lambda} = - j_{LL}^2 - j_{LR}^2 \int_0^{\Lambda} dE \; P(E)\;.
\end{equation}
In the following we analyze this result.

\subsubsection{General considerations}

The flow equations~(\ref{flow1}) and (\ref{flow2}) are the main result of this Section. 
They describe the weak-coupling
behavior of the Kondo model in a generic environment characterized by the $P(E)$
distribution, under the quasi-elastic approximation.
We now specialize to the case of ohmic dissipation, for which $P(E)$ was discussed in 
Sec.~\ref{sec:dynCB}. 

Although the complete form of $P(E)$ is complicated, 
under the assumption that $\w_c\ll U$ (consistent with the quasi-elastic
approximation) we can roughly sketch the
behavior of the flow in a simple two-step picture with respect to the energy 
scale $\w_c$. At the beginning of the flow, $\Lambda\gg\w_c$ and the energy integral 
in Eq.~(\ref{flow2}) is unity since $P(E)$ is a probability. One thus recovers the usual 
Kondo flow equations, i.e. dissipation is ineffective in this case. This can be 
understood by the disappearance of DCB at high energy,
and will give rise to an initial strong renormalization of all Kondo couplings as
if Coulomb blockade was absent.

This flow continues until $\Lambda$ reaches $\w_c$, where the anomalous low-energy 
behavior of $P(E)$ given by Eq.~(\ref{low}) comes into play. At this stage,
it is useful to introduce new dimensionless couplings 
$\lambda_{\al,\be}\equiv \sqrt{A_r/2r} (\Lambda/\w_c)^r j_{\al,\be}$ for
$\al\neq\be$, and $\lambda_{\al,\al} = j_{\al,\al}$, which allow to rewrite the
flow equations as
\begin{subequations}
  \label{JJ}
  \begin{eqnarray}
    \label{JB}
    \frac{d \lambda_{LR}}{d\log \Lambda} & = & r\lambda_{LR} -
    \lambda_{LR}(\lambda_{LL}+\lambda_{RR}) \\
    \label{JF}
    \frac{d \lambda_{LL}}{d\log \Lambda} & = & - \lambda_{LL}^2 - \lambda_{LR}^2\;.
  \end{eqnarray}
\end{subequations}
Let us first give some general comments on these equations. First we notice that
the dissipation affects the inter-lead couplings Eq.~(\ref{JB}) and drives them
irrelevant, at least as long as $\lambda_{LL}+\lambda_{RR}$ is not too
large. This is consistent with the picture where charge transfer across the quantum 
dot is suppressed by DCB, and has the clear
consequence of diminishing the Kondo temperature at increasing values of $r$
(see also the discussion in Sec.~\ref{sec:TK} and corresponding
Fig.~\ref{ll}).
Second, these flow equations are reminiscent of those obtained for the
tunneling through a magnetic impurity in a Luttinger liquid,\cite{fabrizio}
although we did not obtain them from bosonization. This identification will 
be put on rigorous grounds in Sec.~\ref{mapping} by virtue of an exact mapping between the two 
problems. This also allows the computation of transport properties of the 
device at low temperature, that the weak-coupling calculation is unable to
capture.

The above two-step argument thus shows that the slow onset of DCB 
at the scale $\w_c$ reinforces the flow with respect to a pure Luttinger flow of the
type~(\ref{JJ}), so that the decrease of the Kondo temperature with
increasing dissipation will not be as pronounced as in the case of a pure
Luttinger chain. 

We also note that the cutting of $P(E)$ at $\w_c$ is certainly not quantitative since 
$P(E)$ is a broad distribution, and we will solve the flow Eqs.~(\ref{flow1}) and (\ref{flow2}) 
numerivally, with the full evaluation of $P(E)$
according to the integral Eq.~(\ref{inteq}), as performed below.

A final remark concerns the case of strong Coulomb blockade for $r\gg1$. In this
regime, the Coulomb charging energy $E_c=e^2/2C$ emerges as a natural cutoff~\cite{note1} 
with
$P(E)\simeq \delta(E-E_c)$. For this case the two-step argument becomes exact,\cite{SFAR} 
and leads to a strong renormalization of the Kondo temperature.

\subsubsection{Kondo temperature}
\label{sec:TK}

\begin{figure}[ht]
\epsfig{figure=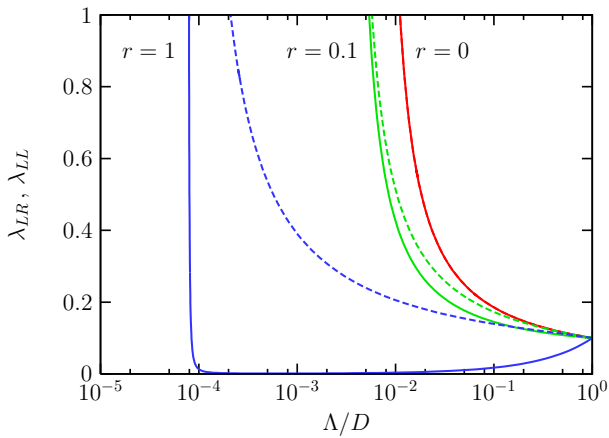,width=8.0cm}
\caption{(Color online) Flow of the Kondo couplings $\lambda_{LR}(\Lambda/D)$ (full line) and $\lambda_{LL}(\Lambda/D)$ 
(dashed line) according to Eqs.~(\ref{JJ}) (i.e. pure Luttinger liquids), 
for different values of $r$. The initial values of the Kondo couplings at scale
$\Lambda=D$ are taken here as $\lambda_{LL}=\lambda_{LR}=0.1$ (with $L/R$
symmetry).}
\label{ll}
\end{figure}

We start for simplicity by considering the solution of the pure Luttinger liquid case,
described by Eqs.~(\ref{JJ}). This corresponds to the regime
of large $\w_c$, ignoring the mixing of boson excitation energies in the
initial Kondo coupling. The results displayed in Fig.~\ref{ll} 
however illustrate in the simplest manner the role of the parameter $r$.
The first general effect is, as expected, a systematic decrease of the Kondo
temperature $T_K$ with increasing $r$, marking a systematic onset of a strong-coupling regime. 
We note that $T_K$ remains nonzero even for strong dissipation, contrarily to the 
noisy "charge" Kondo effect,\cite{karyn1,karyn2}
in which voltage fluctuations act as a fluctuating magnetic field would act in 
the spin Kondo problem, that may prevent the formation of the Kondo resonance.
Apart from this quantitative effect, we infer from Fig.~\ref{ll} 
that dissipation can furthermore discriminate between several strong-coupling fixed points.
Indeed when the impedance $R$ becomes comparable to the quantum value $R_K$, the
inter-lead couplings $\lambda_{LR}$ are strongly driven to zero. This should clearly
lead to qualitatively different transport properties compared to the case of
weak dissipation. This also allows for a stabilization of a two-channel
Kondo effect by DCB, for the case of balanced $\lambda_{LL}$ and
$\lambda_{RR}$ couplings.
These aspects are confirmed by the strong-coupling analysis performed
in Sec.~\ref{mapping}.

\begin{figure}[ht]
\epsfig{figure=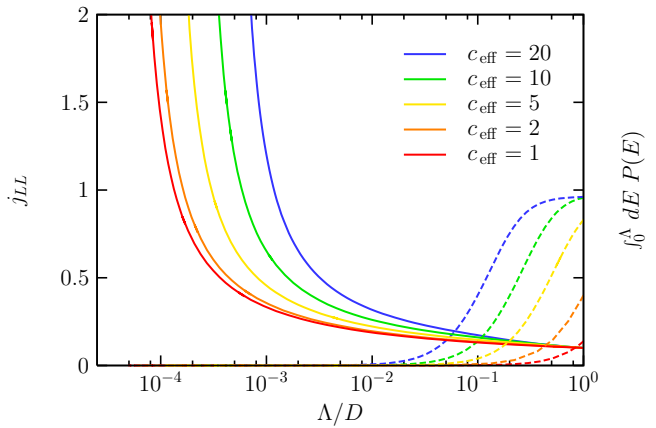,width=8.5cm}
\caption{(Color online) Flow of the Kondo coupling $j_{LL}(\Lambda/D)$ 
(full lines) according to Eqs.~(\ref{flow1}) and (\ref{flow2}), and the corresponding integrated distribution 
$\int_0^\Lambda dE\; P(E)$ (dashed lines), for the dissipation strength $r=1$, initial couplings 
$j_{LL}=j_{LR}=0.1$ 
and several values of the dimensionless effective capacitance $c_{\rm eff}= D R_K
C_{\rm eff}=1,2,5,10,20$ (from bottom to top).}
\label{diss}
\end{figure}

\begin{figure}[ht]
\epsfig{figure=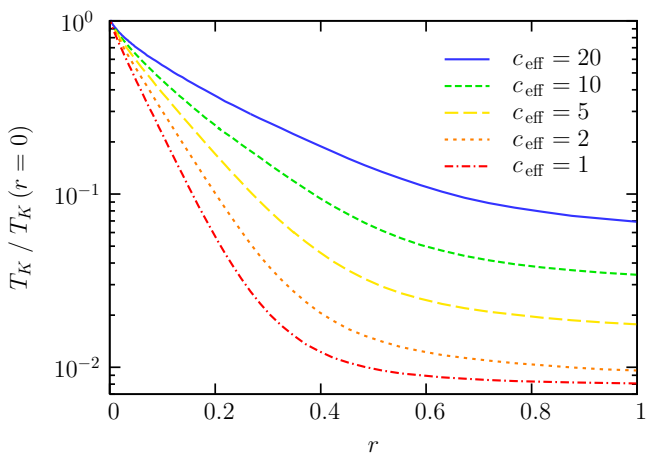,width=8.5cm}
\caption{(Color online) Kondo temperature $T_K$ as a function of dissipation $r$, for several
values of the dimensionless effective capacitance $c_{\rm eff}
=1,2,5,10,20$ (from bottom to top).}
\label{tk_c}
\end{figure}

The solution of the flow Eqs.~(\ref{flow1}) and (\ref{flow2}) with
ohmic dissipation, using the full $P(E)$ function determined by
Eq.~(\ref{inteq}), is shown in Fig.~\ref{diss}. It illustrates the 
general fact that the full distribution function $P(E)$ of the environmental
modes (or equivalently the complete frequency-dependent impedance of the junction 
in the circuit) affects the determination of the Kondo temperature in a significant
quantitative manner.
This is also seen by a systematic study of the Kondo temperature (defined here by the
criterion $j_{LL}(\Lambda\!=\!T_K)=10)$ as a function of dissipation $r$ and for
several values of the effective impedance $C_{\rm eff}$ in Fig.~\ref{tk_c}.

\subsection{Transport anomalies}
\label{mapping}

\subsubsection{Equivalence to the Kondo model between Luttinger liquid leads}

The previous weak-coupling analysis already informs about the
type of ground state favored by the environment. More sophisticated tools are 
however necessary to make precise
statements on the nature of the ground state, and to compute low-temperature 
and low-voltage properties.
Here we will formulate the low-energy equivalence between the
Hamiltonian~(\ref{HK}) and the problem of a $S=1/2$ magnetic 
impurity weakly coupled to two Luttinger liquids, in the tunneling geometry. 
This equivalence has been already shown at the Hamiltonian level for a nonmagnetic 
impurity by Safi and Saleur~\cite{safi} (see also Ref.~[\onlinecite{karyn2}]), and can be 
easily extended to the Kondo model through bosonization of the lead 
electrons,\cite{matveev} as derived in App.~\ref{app}. The final result of
this calculation allows to identify the effective interaction parameter of the Luttinger
chain as
\beq
K = \frac{1}{1+2r}\;.
\eeq
The problem of a magnetic impurity in a Luttinger liquid has been analyzed using
weak-coupling renormalization-group equations and a stability analysis of
the strong-coupling limit by Fabrizio and Gogolin~\cite{fabrizio}, see also
Ref.~[\onlinecite{kim}] for a conformal field theory analysis. It is thus not
surprising that the low-energy limit~(\ref{JJ}) of the
flow Eqs.~(\ref{flow1}) and (\ref{flow2}) reproduces those results.
We emphasize that the flow equations obtained for the alternative "side"
geometry~\cite{leetoner,furusaki} differ from the present "tunneling" geometry
ones by a relevant scale dimension $1>K$ of the inter-lead couplings rather 
than the irrelevant scale dimension $1/K>1$ (as $r>0$) found above. The resulting 
physics is thus markedly different for the two cases.

\subsubsection{Strong-coupling analysis}
The low-energy physics and therefore transport and thermodynamic properties are dominated by
the vicinity of the strong-coupling fixed point. In order to analyze the low-energy physics,
two distinctive regimes arise depending on the value of $r$. 

{\em Small values $r<1/2$.}
The previous weak-coupling analysis has shown that all Kondo interaction 
terms are driven to strong coupling such that the usual one-channel Kondo 
physics should {\it a priori} 
be recovered. However, for a more precise statement 
the vicinity of this fixed point should be considered. 
A detailed analysis shows that particle-hole symmetry-breaking terms
 are relevant at the 
one-channel Kondo fixed point. 
One indeed finds that potential scattering terms like $V c^\dag c$ in Eq.~(\ref{scat})   have a
 scaling dimension  $(1+K)/2=(1+r)/(1+2r)<1$ and are therefore relevant. 
In a one-dimensional infinite Luttinger liquid chain, such a term ``cuts'' the chain
into two semi-infinite chains below an energy scale $T^*$.\cite{kane} 
Therefore, the ultimate fixed point which the system 
reaches for $T=eV=0$ corresponds to an insulating one.\cite{fabrizio,kim,simon01} 
The analysis of the operator content at this insulating fixed point allows to
predict the scaling behavior with temperature or bias voltage.
Here we focus on the differential conductance
$dI/dV$ between the left and right lead.
The leading irrelevant operator corresponds to a hopping term between the two leads 
and has dimension $(1+1/K)/2=1+r>1$. 
We note that particle-hole symmetry-breaking 
effects can be tuned to zero, using the dot gate voltage. 
The Kondo resonance may therefore develop in transport 
for some particular value of the gate voltage $V_g^*$, as specified in
Eq.~(\ref{G2}) below. 
The finite-temperature corrections to the unitary limit are given by the leading irrelevant operator nearby this perfectly transmitting fixed point when particle-hole symmetry is imposed. This operator has dimension $2K$ and is irrelevant for $K>1/2$.\cite{kim}
These results imply an anomalous behavior of the differential conductance
at low temperature:
\begin{subequations}
\label{G12}
 \begin{eqnarray}
\label{G1}
\frac{dI}{dV} &\approx& a \left(\frac{eV}{T^*}\right)^{2r}~{\rm for}~V_g\neq V_g^*\\
\label{G2}
\frac{dI}{dV} &\approx& 2 G_K\left[1-b \left(\frac{eV}{T_{K}}\right)^{2r}\right]~{\rm for}~V_g=V_g^* 
\end{eqnarray}
\end{subequations} 
with $G_K=1/R_K=e^2/h$, and $a,b$ dimensionless nonuniversal prefactors. 
At finite temperature $T$ and zero bias voltage, the energy $eV$ should be replaced by $T$ in Eqs.~(\ref{G12}).
The conductance in Eq.~(\ref{G1}) is similar to the conductance through a 
single tunnel junction. One may also wonder how 
the conductance is modified when $V_g$ slightly deviates
from $V_g^*$. Since particle-hole symmetry is relevant at this perfectly
transmitting fixed point, and the resonance is very sharp. Following Ref.
[\onlinecite{kane}], it can be shown that the width of the resonance scales as
$T^{(1-K)/2}$ where $V_g$ deviates from $V_g^*$. The differential conductance
escapes from the unitary limit as follows
\beq
\frac{dI}{dV} \approx 2 G_K\left[1-b \left(\frac{\Gamma}{eV}\right)^{\frac{2r}{1+2r}}\right]~{\rm for}~V_g\approx V_g^* 
\eeq
where $\Gamma$ is a characteristic energy scale related to $\delta V_g=V_g-V_g^*$, which vanishes as
$\delta V_g^{2/(1-K)}$ at resonance. \cite{kane}

{\em Large values $r>1/2$.}
For large values of $r$, dissipation dominates the renormalization-group flow in 
Eq. (\ref{JB}) and the Kondo coupling $J_{LR}$ renormalizes to zero. 
Therefore coherent charge transfer between left and right lead is suppressed, without however 
completely inhibiting the Kondo effect, similarly to what occurs in the case of
strongly Coulomb blockaded leads~\cite{oreggold,SFAR}. Indeed the Kondo couplings $J_{LL}$ and $J_{RR}$ 
are still driven to strong coupling according to Eq.~(\ref{JF}).
A two-channel Kondo fixed point can be reached provided $J_{LL}=J_{RR}$. 
By analyzing the operator content of the strong-coupling two-channel 
fixed point, it has been shown to be stable against particle-hole symmetry-breaking terms
for $K<1/2$, or equivalently $r>1/2$. \cite{fabrizio,kim}
Indeed, the leading irrelevant operator near the two-channel Kondo fixed point 
occurs in the flavor sector and has dimension $1/(2K)=2r+1$,
and is therefore irrelevant for $r>1/2$.\cite{kim}
This operator will therefore dominate the  transport properties below the two-channel Kondo temperature $T_{2CK}$.
For $J_{LL}\neq J_{RR}$ the two-channel Kondo fixed point is unstable and a single-channel Kondo
effect occurs in the strongest coupled channel (either left or right). 
The transport for this situation is 
characterized by the leading irrelevant operator corresponding
to a hopping term between the two leads as for $r<1/2$ with dimension $(1+1/K)/2=1+r>1$.
From this above analysis we can infer the anomalous behavior of the differential conductance
$dI/dV$ between the left and right leads at low energy:
\begin{subequations}
\bea
\label{G3}
\frac{dI}{dV} &\approx& c \left(\frac{eV}{T_{1CK}}\right)^{2r}~{\rm for}~J_{LL}\neq J_{RR},\\
\label{G4}
\frac{dI}{dV} &\approx& d \left(\frac{eV}{T_{2CK}}\right)^{2r-1}~{\rm for}~J_{LL}=J_{RR}\;,
\eea
\end{subequations}
where $c,d$ are nonuniversal numbers and $T_{1CK}$ is the one-channel (eithr left or right) 
Kondo temperature. 
Notice the absence of $G_K$, peculiar to the two-channel Kondo effect 
in Eq.~(\ref{G4}), due to the tunneling geometry. A third, nonnoisy
electrode would be necessary to obtain this behavior, see
Ref.~[\onlinecite{oreggold,SFAR}].

This analysis can be directly extended to a quantum dot operating 
as a spin-$1$ impurity. The low-energy physics resembles then that of a 
spin-$1$ impurity embedded between two Luttinger liquids, 
recently studied in Ref.~[\onlinecite{durga}].

\section{Conclusion}

In this paper we investigated the impact of environmental Coulomb 
blockade on the Kondo effect in the case of moderate to highly resistive leads. 
An ohmic resistance of the environment exceeding $R_K/2$
induces a suppression of the inter-lead Kondo interactions,
without however preventing the formation of a strong-coupling state, due 
to the remaining intra-lead processes. When the tunneling amplitudes between the left and
right leads are equal, even a two-channel Kondo effect can be reached.
For an environmental resistance smaller than $R_K/2$ the Kondo effect can fully develop
between both leads. However, the fully transparent fixed point is stable only when
particle-hole symmetry is maintained with the dot plunger gate voltage. 
On a qualitative level these results imply anomalous low-temperature transport 
properties through the device, even though the Kondo effect survives. More quantitatively,
the Kondo temperature below which those features start to appear was shown to
depend sensitively not only on the dissipation strength, but rather on the full
spectral function of the excited environmental modes. Although strong-resistive
effects may be difficult to detect in actual experiments, due to the difficulty in
the realization of well-coupled highly dissipative setups in semiconducting quantum
dots, such a reduction of the Kondo temperature at moderate dissipation 
as well as nonlinearities in the I(V) characteristics
should be observed.

Theoretically, it would be interesting to go beyond the quasi-elastic approximation 
by which our phenomenological Kondo model is applicable. This approximation
may rather underestimate the environmental effects.
A generalization
of this study to orbital Kondo effects or mixed spin-orbital Kondo setups 
can also be performed along the same line~\cite{doubledot}
merging the present situation with the previously studied
decoherence effects on the Kondo effect in the charge sector.

\acknowledgments{This work has been supported by the French ANR (program PNano),
the DFG Center for Functional Nanostructures in Karlsruhe, and the Virtual Quantum 
Phase Transitions Institute in Karlsruhe.
We would also like to acknowledge valuable discussions with K.~Le Hur at the early
stages of this work.}

\appendix

\section{Equivalence between the noisy Kondo problem and the Kondo model in
Luttinger liquids}
\label{app}

We provide here the derivation for the low-energy mapping of the Kondo
model~(\ref{HK}) in an ohmic environment onto the usual Kondo model between
Luttinger liquids.

The free electronic part of Eq.~(\ref{leads}) is first conveniently expressed as a
chiral Hamiltonian for left and right electrodes
\begin{eqnarray}
\label{chiral}
H_{\rm leads} & = & -iv_F \intx \; \sum_{\s\al} c^\dag_{\s,\al} \partial_x c^\pdag_{\s,\al}\;.
\end{eqnarray}
We bosonize each channel, keeping the Klein factors explicitely
\begin{equation}
\label{boso}
c^\pdag_{\s\al} = \frac{1}{\sqrt{2\pi a}} F_{\s\al}
\exp\left(i\frac{\phi_{c\al}+(-1)^\s \phi_{s\al}}{\sqrt{2}}\right)\;.
\end{equation}
This allows to express the above Eq.~(\ref{chiral}) in terms of charge and
spin bosons in each channel
\begin{eqnarray}
\label{free}
H_{\rm leads} & = & \frac{v_F}{4\pi} \intx \; \sum_{\al} 
(\partial_x\phi_{c\al})^2+ (\partial_x\phi_{s\al})^2\;.
\end{eqnarray}
Using the identity for the electron density operator
\begin{equation}
c^\dag_{\s\al} c^\pdag_{\s\al} = \frac{1}{2\pi}
\frac{\partial_x \phi_{c\al} + (-1)^\s \partial_x \phi_{s\al} }{\sqrt{2}}
\end{equation}
and the bosonization dictionary~(\ref{boso}), we can now re-express the Kondo
Hamiltonian~(\ref{HK}). To clarify the resulting expression, we define the 
following basis change between the $L,R$ channels 
\begin{subequations}
\begin{eqnarray}
\phi_c & \equiv & \frac{\phi_{cL} + \phi_{cR}}{\sqrt{2}} \\
\phi_f & \equiv & \frac{\phi_{cL} - \phi_{cR}}{\sqrt{2}} \\
\phi_s & \equiv & \frac{\phi_{sL} + \phi_{sR}}{\sqrt{2}} \\
\phi_{sf} & \equiv & \frac{\phi_{sL} - \phi_{sR}}{\sqrt{2}}\;.
\end{eqnarray}
\end{subequations}
We have also to perform a similar transformation on the Klein factors~\cite{ZvD}
and introduce for this purpose four alternate Klein fermions, $F_c$, $F_f$, $F_s$, $F_{sf}$, 
which obey
\begin{subequations}
\begin{eqnarray}
F^\dag_{sf} F^\dag_{s} &= &F^\dag_{e\uparrow} F^\pdag_{e\downarrow} \\
F^\pdag_{sf} F^\dag_{s}& = &F^\dag_{2\uparrow} F^\pdag_{2\downarrow} \\
F^\dag_{sf} F^\dag_{f} &= &F^\dag_{e\uparrow} F^\pdag_{2\uparrow} \\
F^\dag_{c} F^\pdag_{s}& =&F^\dag_{e\uparrow} F^\dag_{2\uparrow}\;.
\end{eqnarray}
\end{subequations}
All other bilinears of the original Klein factors are easily computed from the
above expression, e.g. 
$ F^\dag_{e\downarrow} F^\pdag_{2\uparrow} = 
F^\dag_{e\downarrow} F^\pdag_{e\uparrow} F^\dag_{e\uparrow} F^\pdag_{2\uparrow} = 
F^\pdag_{s} F^\pdag_{sf} F^\dag_{sf} F^\dag_{f} = 
F^\pdag_{s} F^\dag_{f}$.

We introduce also spin anisotropies of the exchange constants, with the
new notation ($J_F^\perp$, $J_F^z$, $J_B^\perp$, $J_B^z$), where the subscript $F$ ($B$) 
stands for $J_{LL}=J_{RR}$ ($J_{LR}=J_{RL}$), and the
superscript $\perp$ ($z$) for the $x,y$ ($z$) spin
orientations respectively. We assumed here perfect symmetry between the left and right
couplings not to overburden the notation, although this will not affect
the validity of the mapping we want to establish.
The final expression for the Kondo term~(\ref{HK}) then reads
\begin{eqnarray}
\nonumber
H_K & = & \frac{J_F^\perp}{4\pi a} S^+ F^\pdag_s e^{i\phi_s}
\left( F^\pdag_{sf} e^{i\phi_{sf}} + F^\dag_{sf} e^{-i\phi_{sf}}\right) + \rm{H.c.}\\
\nonumber
& + & \frac{J_F^z}{2\pi} S^z \partial_x \phi_s \\
\nonumber
& + & \frac{J_B^\perp}{4\pi a} S^+ F^\pdag_s e^{i\phi_s}
\left( F^\pdag_{f} e^{i\phi_{f}+i\vp} + 
F^\dag_{f} e^{-i\phi_{f}-i\vp} \right) + \rm{H.c.}\\
\nonumber
& + & \frac{J_B^z}{4\pi a} S^z F^\dag_{f} e^{-i\phi_{f}-i\vp}
\left( F^\dag_{sf} e^{-i\phi_{sf}} - F^\pdag_{sf} e^{i\phi_{sf}}\right) + \rm{H.c.} \;.\\
\end{eqnarray}

We see that the environmental boson $\vp$ occurs naturally as a systematic shift
of the "flavor" field $\phi_f$ describing the charge transfer
between the two electrodes. To define a simple effective environment, we
consider the combined effective action for both the local density bosons
(integrating all $x\neq0$ in Eq.~(\ref{free})), and the environmental boson
from Eq.~(\ref{spectrum})
\begin{eqnarray}
\nonumber
S_0 & = & \frac{1}{\beta} \sum_{\nu_n} |\nu_n| \left( |\phi_c|^2 + 
|\phi_f|^2 + |\phi_s|^2 + |\phi_{sf}|^2 + \frac{R_K}{2R} |\vp|^2 \right) \\
\nonumber
& = & \frac{1}{\beta} \sum_{\nu_n} |\nu_n| \left( |\phi_c|^2 + 
|\phi_f'|^2 + |\phi_s|^2 + |\phi_{sf}|^2 + |\vp'|^2 \right) \;,
\end{eqnarray}
where
\begin{eqnarray}
\phi_f' & = & \sqrt{K} \left(\phi_f+\vp\right)\\
\vp' & = & \sqrt{K} \left( \sqrt{\frac{2R}{R_K}}\phi_f-
\sqrt{\frac{R_K}{2R}}\vp\right)
\end{eqnarray}
and
\begin{eqnarray}
K & \equiv & \frac{1}{1+2R/R_K} \;.
\end{eqnarray}\\
Therefore the role of the ohmic environment on the Kondo effect is to replace
all exponential terms $e^{i\phi_f}$ in the bosonized expression by 
$e^{i\phi_f'/\sqrt{K}}$. Starting from a standard magnetic impurity between
two Luttinger liquid leads characterized by an interaction parameter $K$
(without environmental modes), with the open boundary bosonization
condition,\cite{fabrizio} leads to the same final expression, thus proving the 
desired equivalence.

\end{document}